# Application of the Cox Regression Model for analysis of Railway Safety Performance

Hendrik Schäbe (TÜV Rheinland) & Jens Braband (Siemens Mobility)

**Abstract**

The assessment of in-service safety performance is an important task, not only in railways. For example it is important to identify deviations early, in particular possible deterioration of safety performance, so that corrective actions can be applied early. On the other hand the assessment should be fair and objective and rely on sound and proven statistical methods.

A popular means for this task is trend analysis. This paper defines a model for trend analysis and compares different approaches, e. g. classical and Bayes approaches, on real data. The examples show that in particular for small sample sizes, e. g. when railway operators shall be assessed, the Bayesian prior may influence the results significantly.

**Introduction**

The analysis of railway safety performance has been of interest for many years. On a nationwide level already so called Common Safety Targets and Common Safety Indicators have been proposed (EU (2009)). Especially, it is important to analyse trends and to derive judgement on whether the performance is improving, deteriorating or unchanged. This is performed today by the European Union Agency for Railways mainly based on weighted averages or moving averages (European Union Agency for Railways (2018)). Due to the normally sparse data, in many cases no judgement can be made, since the small sample size does not allow to provide a statistically significant result.

In this paper we analyse numbers of severe railway accidents as reported by Evans (2020). Note that this topic is of particular relevance as the European Railway Agency is drafting a legal text on the assessment of safety levels and safety performance of railway operators in the EU. But in the current draft (European Union Agency for Railways (2020)) the term safety level is used for the statistical safety performance and is defined currently in a very narrow sense: "'safety level' means the weighted sum of occurrences of eligible events … corresponding to a given volume of operation … and normalised by this volume of operation…". Currently also no statistical procedure for the evaluation of these safety levels has been proposed so that this paper tries to fill the gap.

For this purpose already several proposals have been made e. g. Evans (2020) or Andrasik (2020), which are taken into account and compared.

**The model**

Railway accidents, in particular severe accidents with fatalities, are rare events and as such the number of severe accidents, say K, follow a Poisson distribution, see Braband and Schäbe (2014):

$$P(K=k) = \frac{\lambda^k}{k!} \exp(-\lambda). \tag{1}$$

If the accidents relate to n different time intervals, which are numbered by the index i, this easily yields

$$P(K_i=k_i) = \frac{\lambda_i^{k_i}}{k_i!} \exp(-\lambda_i). \tag{2}$$

For different time intervals, we have here assumed different intensities $\lambda_i$, since we may not assume a priori, that the intensity with which the accidents occur is always the same.

Often the $\lambda_i$ are expected to form a decreasing sequence of values. In order to model this effect, we use the Cox regression model using the time as explanatory variable, see Cox and Oakes (1984).

Then we have:

$\lambda_i = \lambda_0 \exp(-\beta t_i)$. (3)

Practically (3) means that we assume n equidistant time intervals, where $t_i$ is the value describing the center of the i-th time interval. Moreover, $\lambda_0$ is the initial intensity of the process of severe accidents. Should the time intervals not be proportional, an additional factor reflecting the different length of the time intervals can easily be introduced. Mainly, one will compare years or five-year intervals with accident data.

Now, the likelihood function is

Lik = $\prod_{i=1}^{n} \frac{\lambda_i^{k_i}}{k_i!} \exp(-\lambda_i)$. (4)

This yields the maximum likelihood estimators, which are obtained by the following equations.

$\hat{\lambda}_0 = \frac{\sum_{i=1}^{n} k_i}{\sum_{i=1}^{n} \exp(-\hat{\beta} t_i)}$ (5)

and

$\sum_{i=1}^{n} k_i t_i \sum_{i=1}^{n} \exp(-\hat{\beta} t_i) = \sum_{i=1}^{n} k_i \sum_{i=1}^{n} t_i \exp(-\hat{\beta} t_i)$ (6)

Note that the estimator for β must be obtained by numerical solution of equation (6). In the next step $\lambda_0$ can be estimated by (5).

Now it is of interest, whether the parameter β is larger than 0 or not. If this is the case, one could conclude a decreasing behavior of the severe accidents.

The fact that a maximum likelihood estimator is asymptotically normally distributed with variance 1/I can be used. Here I is the Fisher information, computed by

I = $-\frac{\partial^2}{\partial \beta^2} \ln(Lik) = \sum_{1}^{n} \lambda_0 t_i^2 \exp(-\beta t_i)$. (7)

Then, a confidence interval for b is

$[\hat{\beta} - z_\alpha \sigma; \hat{\beta} + z_\alpha \sigma]$. (8)

with

$\sigma = \sqrt{I}$ (9)

and $z_\alpha$ is the 1-α Quantile of the Normal distribution. Then, the confidence interval has a coverage of 1-2*α.

The result (8) can also be used for a statistical test for the Hypothesis: there is no decrease in the number of severe accidents. If

UConf = $\hat{\beta} - z_\alpha \sigma$ (10)

is larger than zero, the hypothesis would would be declined and improvement, i.e. a significant decrease of severe accidents over time would be concluded.

It has to be noted that the maximum likelihood estimators are asymptotically efficient, i.e. have the smallest spread among all estimators if the sample size is large enough.

**Examples**

Using the data of Evans (2020) we have analysed the data of several countries.

The figures below show the counts and the fitted curve. Also the last two points of the counts represent the counts for 1980-2019 and 1990-2019 for a 5 years period – for comparison with the other data.

We have used data from the entire European union plus Switzerland plus Norway and also the date from several large countries as Germany, France, Italy and the United Kingdom. As an example for a small country we have used Estonia. Here it becomes evident, that small sample sizes are a real problem.

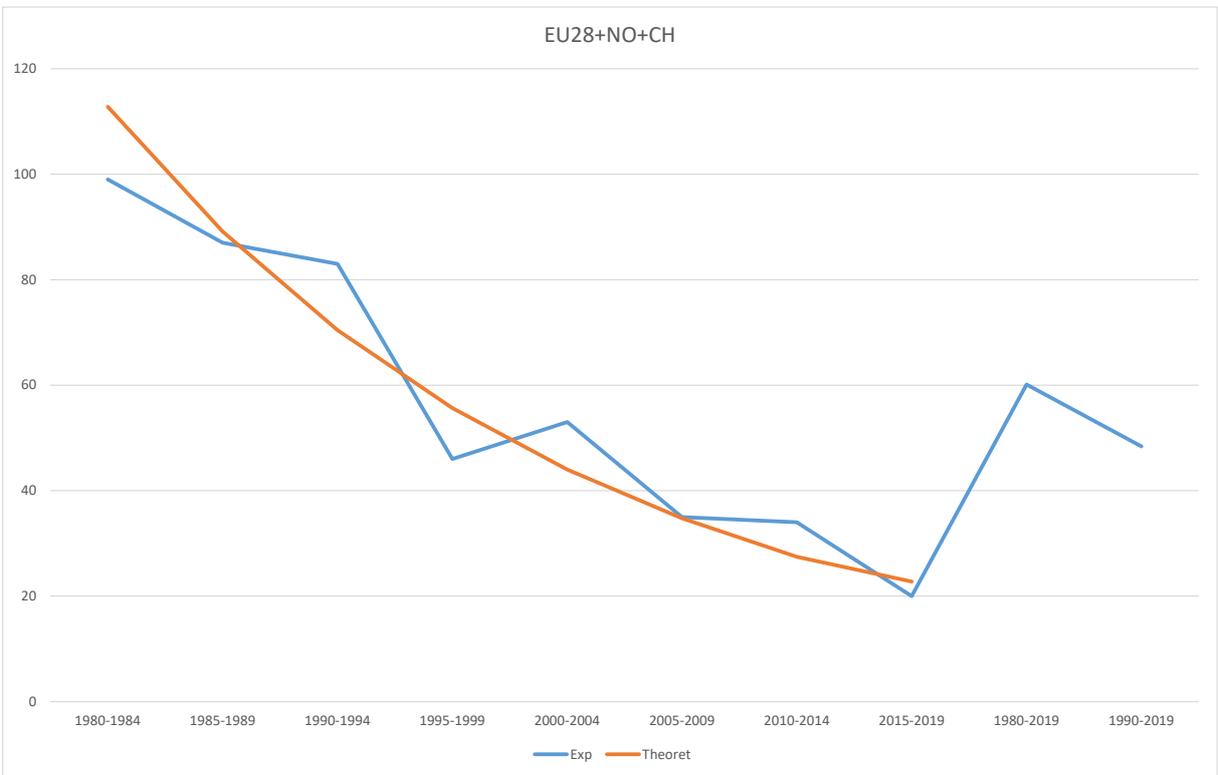

Figure 1       Number of accidents with fatalities (counts and model) of the 28 EU countries + Switzerland + Norway

Since UConf takes the value of 0,043 with first kind error of 10% we conclude a significant decrease.

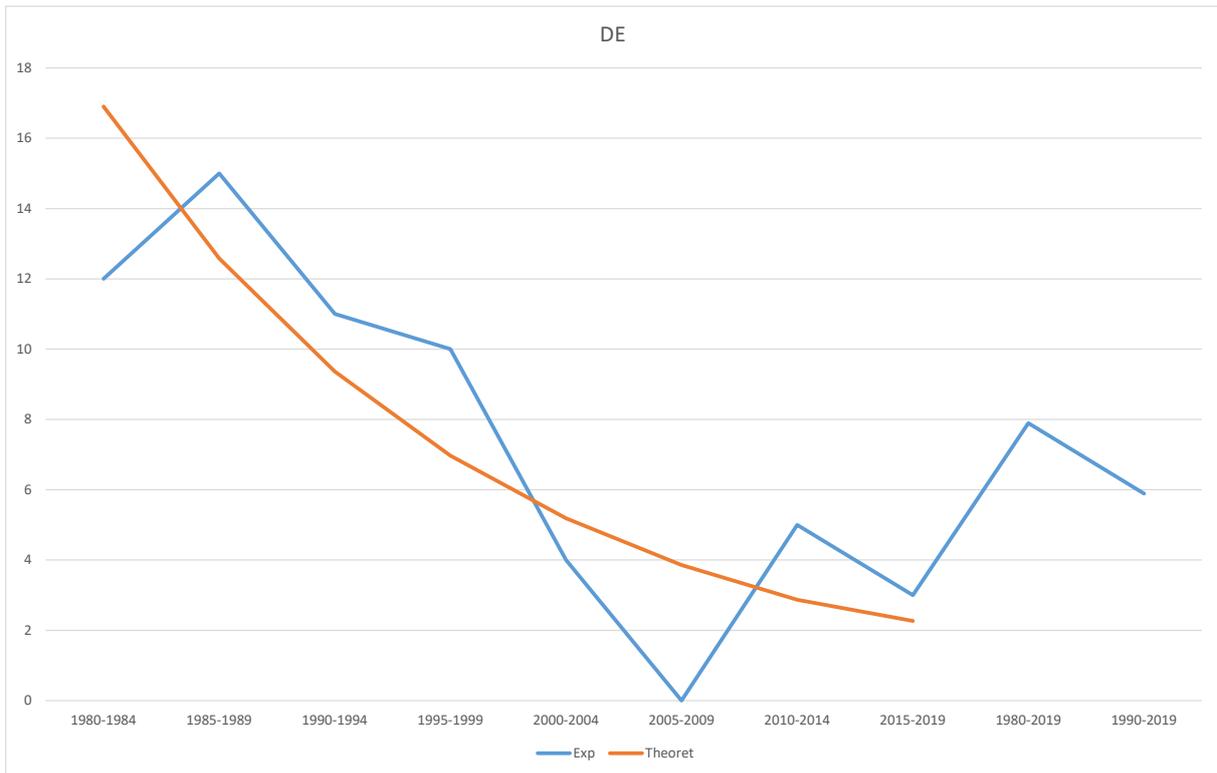

Figure 2	Number of accidents with fatalities (counts and model) of Germany

Since UConf takes the value of 0,046 with first kind error of 10% we conclude a significant decrease.

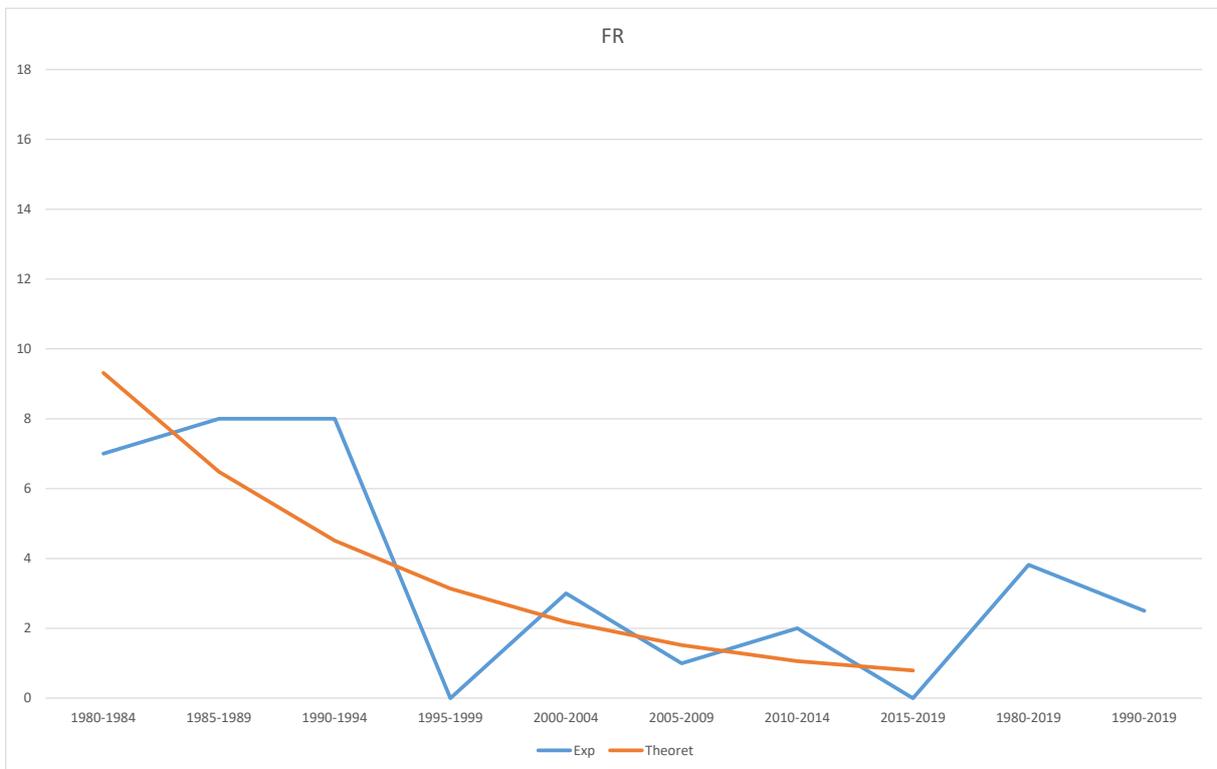

Figure 3	Number of accidents with fatalities (counts and model) of France

Since UConf takes the value of 0,0517 with first kind error of 10% we conclude a significant decrease.

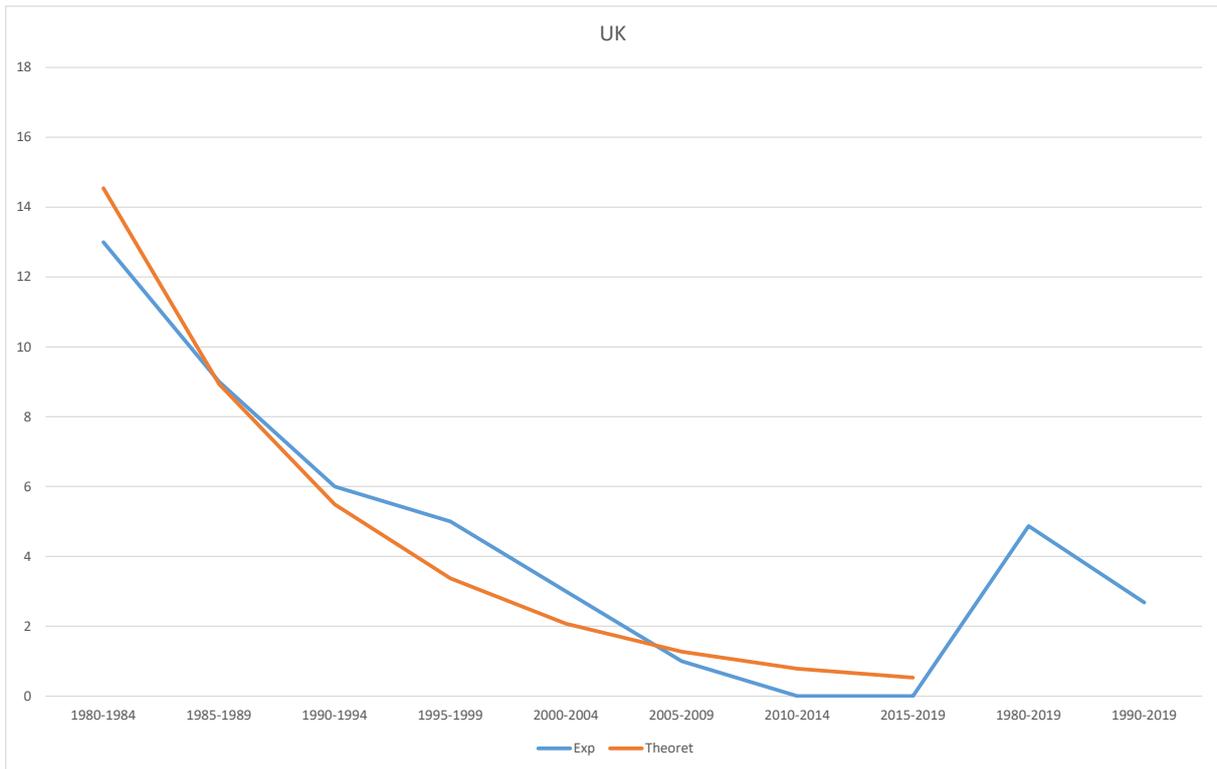

Figure 4    Number of accidents with fatalities (counts and model) of the United Kingdom

Since UConf takes the value of 0,0757 with first kind error of 10% we conclude a significant decrease.

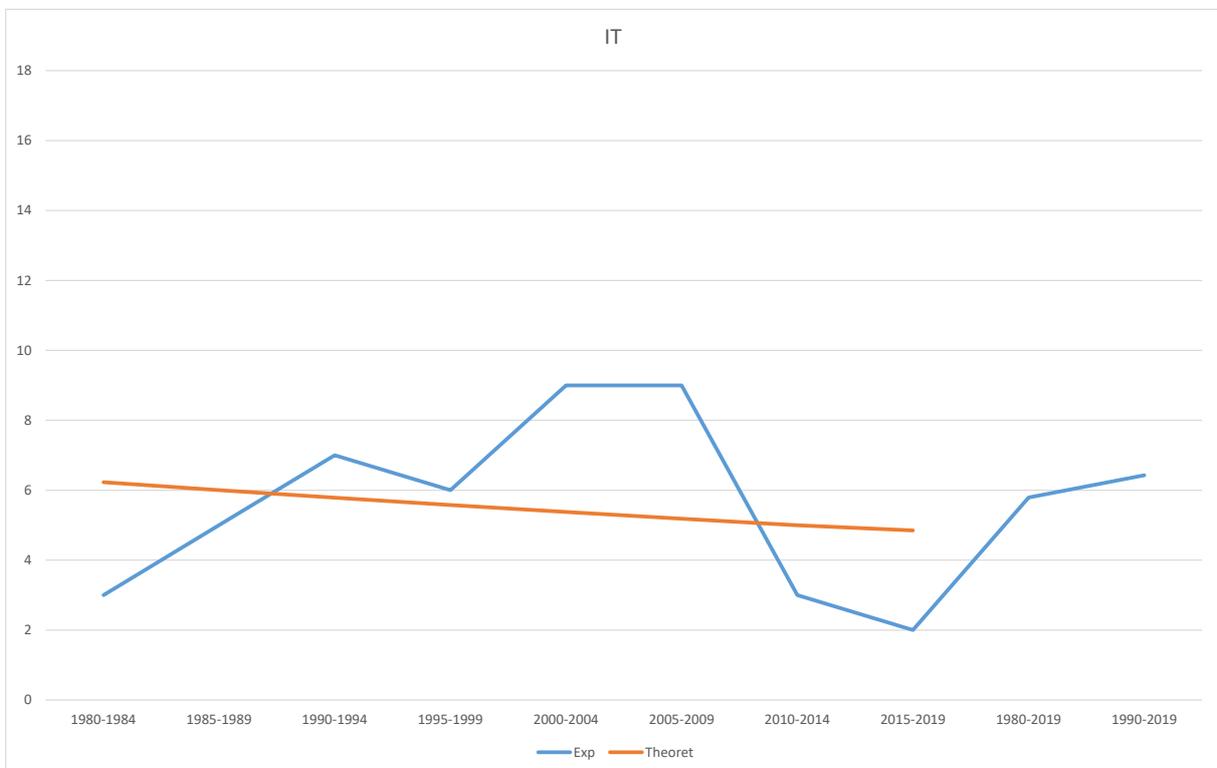

Figure 5    Number of accidents with fatalities (counts and model) of Italy

Since UConf takes the value of -0,00403 with first kind error of 10% we cannot conclude a significant decrease, although visually a decreasing tendency can be seen. This, however, is not significant and can be caused by random influences.

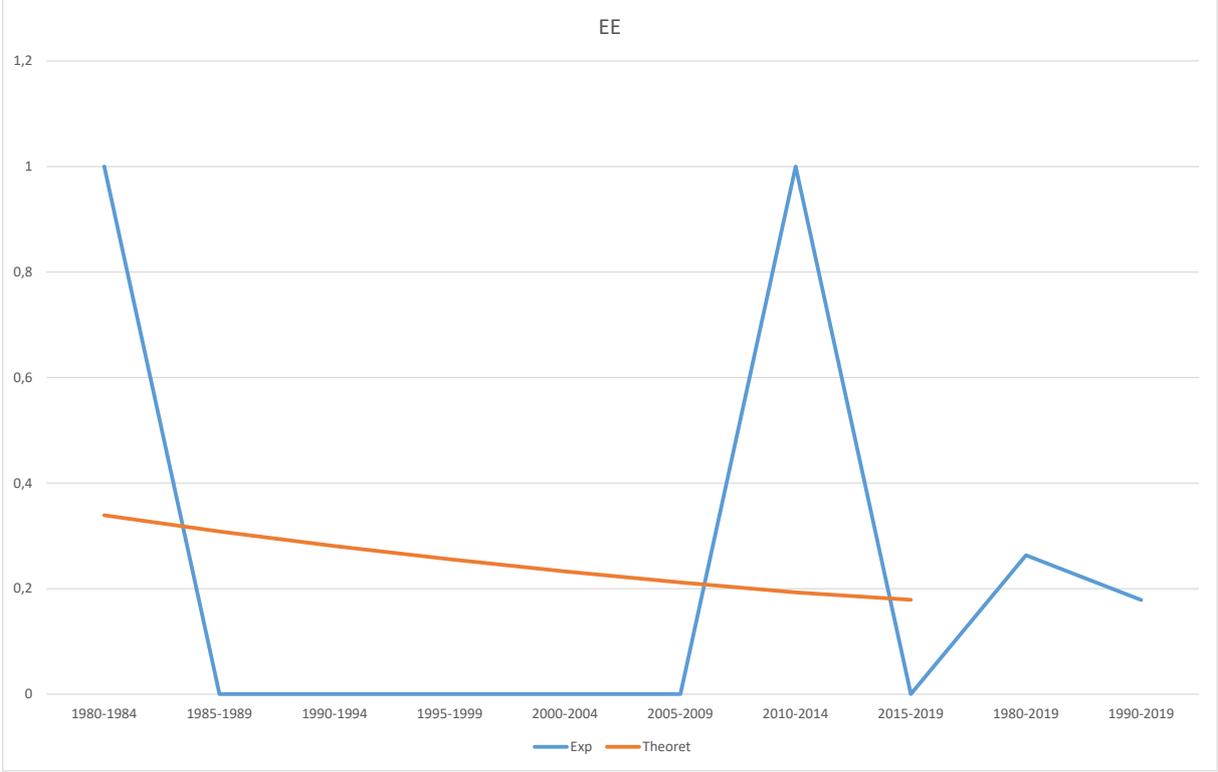

Figure 6          Number of accidents with fatalities (counts and model) of Estonia

Since UConf takes the value of -0,38 with first kind error of 10% we cannot conclude a significant decrease. This, however is a result of the small number of accidents (only two). A general decreasing trend can be seen.

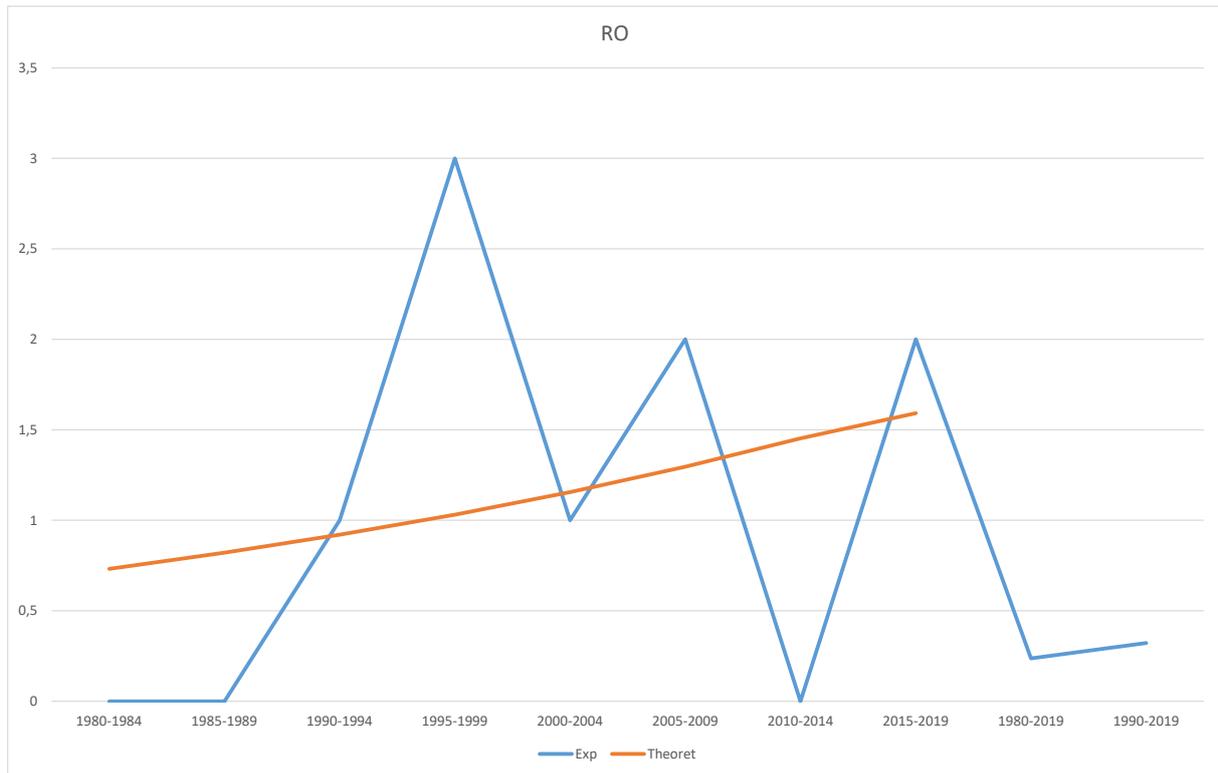

Figure 7    Number of accidents with fatalities (counts and model) of Romania

Since OConf takes the value of -0,00094 with first kind error of 10% we cannot conclude a significant increase, which also visually can be seen

The results how, that the model allows a statistical analysis and a significance test if enough data are available. We have analysed the data collected from 1980 until 2019, which is a long time. This allowed us to derive results in many cases, but not in all.

**Bayesian approach**

Frequently, expert or prior information shall be used. In order to avoid unnecessary complexity, we use a conjugate prior $\pi(\lambda_0,\beta)$ in the same form as the likelihood function (4) with (3), i.,e.

$$\ln(\pi(\lambda_0,\beta)) \sim \sum_{i=1}^{n}[a_i(\ln(\lambda_0) - \beta\tau_i) - \lambda_0\exp(-\beta\tau_i)]. \tag{11}$$

This prior has the effect, as if in the time intervals with central points $\tau_i$ $a_i$ severe accidents would have been occurred. If all $a_i$ would be 0, we would have a non-informative prior.

Due to the use of the conjugate prior, for $t_i = \tau_i$ a Bayes estimator for the maximum of the posterior density is equivalent to the maximum likelihood estimator with data $qn_i+(1-q)a_i$ using (5) and (6). Here qi is the weight that we give to the prior information. An approximate HPD interval (high probability interval) of the posterior density can be obtained from (8) using data $n_i+a_i$. Here, we have approximated the posterior by a normal distribution around it maximal value. One has to note that within this model the information from the prior distribution and from the sample are combined and that leads to a seemingly higher sample size and therefore, to larger values of $\lambda_0$. With one exception- if a non-informative prior is used.

If, one is only interested in the parameter β, for deriving a judgement on an increase or decrease of the number of fatalities on this would not play a role.

**Statistics with different time interval lengths**

The model can easily be adopted for the case of time intervals with different length. Then, the formulae (2) – (6) would be rewritten in the following manner

The Poisson distribution is kept unchanged:

$$P(K_i = k_i) = \frac{\lambda_i^{k_i}}{k_i!} \exp(-\lambda_i). \tag{12}$$

The different intensities $\lambda_i$ now also depend on the lengths of the time intervals, say $T_i$.

Then (3) is changed into:

$$\lambda_i = \lambda_0 \, T_i \exp(-\beta t_i). \tag{13}$$

That means, the intensity is proportional to the length of the interval and to the well-known regression factor and the $\lambda_0$, which is now a parameter per time unit, which is different from the model in (1) – (4). The time $t_i$ lies in the center of the i-th interval having length $T_i$.

Now, the log likelihood function is

$$l = \ln(\text{lik}) = \sum_{i=1}^{n} k_i (\ln \lambda_0 - \beta t_i + \ln(T_i)) - \ln k_i! - \lambda_0 T_i \exp(-\beta t_i) \tag{14}$$

This yields again the maximum likelihood estimators, which are obtained by the following equations.

$$\hat{\lambda}_0 = \frac{\sum_{i=1}^{n} k_i}{\sum_{i=1}^{n} T_i \exp(-\hat{\beta} t_i)} \tag{15}$$

and

$$\sum_{i=1}^{n} k_i t_i \sum_{i=1}^{n} T_i \exp(-\hat{\beta} t_i) = \sum_{i=1}^{n} k_i \sum_{i=1}^{n} t_i \, T_i \exp(-\hat{\beta} t_i) \tag{16}$$

Formula (7) becomes

$$I = -\frac{\partial^2}{\partial \beta^2} \ln(Lik) = \sum_{1}^{n} \lambda_0 T_i t_i^2 \exp(-\beta t_i). \tag{17}$$

For simplicity assume now only two time intervals with $T_1 = \alpha T$ and $T_2 = (1-\alpha) T$ so that $T_1/T_2 = \alpha / (1-\alpha)$ and $T = T_1 + T_2$ and $k = k_1 + k_2$.

Then, $t_1 = \alpha T/2$ and $t_2 = (1+\alpha)T/2$

$$\hat{\lambda}_0 = \frac{k}{\alpha \, T \exp\left(-\frac{\hat{\beta} \alpha T}{2}\right) + (1-\alpha) T \exp\left(-\frac{\hat{\beta}(1+\alpha)T}{2}\right)} \tag{18}$$

$$[k_1\alpha T + k_2(1+\alpha)T]\left[\alpha\, T\exp\left(-\frac{\hat{\beta}\alpha T}{2}\right) + (1-\alpha)\, T\exp\left(-\frac{\hat{\beta}(1+\alpha)T}{2}\right)\right]$$

$$= k\left[\alpha^2 T^2 \exp\left(-\frac{\hat{\beta}\alpha T}{2}\right) + (1-\alpha^2)T^2\exp\left(-\frac{\hat{\beta}(1+\alpha)T}{2}\right)\right] \quad (19)$$

$$I = \lambda_0 T^3\{\alpha^3\exp(-\beta\alpha T/2) + (1+\alpha)(1-\alpha^2)\exp(-\beta(1+\alpha)T/2)\}/4 \quad (20)$$

Using a Bayesian prior distribution with the same $\alpha$ and T but parameters $a_1$ and $a_2$, (18), (19) and (20) keep valid, only with

$k_1$ replaced by $k_1 + a_1$ and

$k_2$ replaced by $k_2 + a_2$.

Again, one need to note that for a statistical inference on the parameter l0 the effect of the Bayes estimator here is that of an increased number of severe accidents.

**Two sample comparison – an example**

In this section we will provide an example for a comparison of the safety behavior of national railway systems in 2010-2014 compared with 2005-2009behavior three cases: Greece, Estonia, European Union. We will use the data from Evans (2020). We will use the Bayesian methods given in the last section. Two different priors will be studied. The first one is a non-informative one with $a_1=a_2=0$. In the second case, we use a prior proposed by Andrasik. Andrasik used a Binomial distribution to model the fatalities in the both time intervals. For intervals of equal length he arrives at parameters $a_1 = a_2 = 2$ in our notation.

Formulae (18) – (20) give in our special case with $\alpha = ½$ and T = 10 years

$$\hat{\lambda}_0 = \frac{2k}{T\left(\exp\left(-\frac{\hat{\beta}T}{4}\right) + \exp\left(-\frac{3\hat{\beta}3T}{4}\right)\right)} \quad (21)$$

$$(k_1 + 3k_2)\left[\exp\left(-\frac{\hat{\beta}T}{4}\right) + \exp\left(-\frac{3\hat{\beta}T}{4}\right)\right]$$

$$= k\left[\exp\left(-\frac{\hat{\beta}T}{4}\right) + 3\exp\left(-\frac{3\hat{\beta}T}{4}\right)\right] \quad (22)$$

This equation can be simplified to give

$$\hat{\beta} = -\frac{2}{T}\ln\left(\frac{z-1}{3-z}\right) \text{ with } z = (k_1+3k_2)/k = 1 + 2k_1/k \quad (23)$$

The information becomes

$$I = \lambda_0 T^3\{\exp(-\beta T/4) + 9\exp(-3\beta T/4)\}/32 \quad (24)$$

As a result we get

$\lambda_1 = k_1$ \hfill (25)

$\lambda_2 = k_2$, \hfill (26)

which is a natural result, since we have two parameters and two values.

The Bayesian point estimators are then

$$\lambda_1 = k_1 + a_1$$

$$\lambda_2 = k_2 + a_2.$$

This result documents two influences. First, the priori distribution augments additional events, so that the estimator give larger values than the classical one. This is caused by the fact that we do not estimate here intensive parameters (not depending on the sample size), but absolute values (extensive parameters). Furthermore, we estimate two parameters from a sample consisting of two elements so that the formula has the same simple form as (25) and (26)

In a next step we apply the algebra to the data of Evans (2020) to derive a judgement on whether the safety behavior is improving or deterioration. This is judged from the parameter $\beta$. The result is given in the following table.

|  |  |  | Classical Test |  |  |  | Bayes Test |  |  |  |
|---|---|---|---|---|---|---|---|---|---|---|
|  | 2005-2009 | 2010-2014 | Information I | Beta (estimator) | lower confidence level | upper confidence level | Information I | Beta (estimator) | lower confidence level | upper confidence level |
| EU28+NO+CH | 35 | 34 | 431,25 | 5,798E-03 | -7,341E-02 | 8,500E-02 | 456,25 | 5,480E-03 | -7,153E-02 | 8,249E-02 |
| Greece (EL) | 2 | 1 | 18,75 | 1,386E-01 | -2,412E-01 | 5,185E-01 | 43,75 | 5,754E-02 | -1,911E-01 | 3,062E-01 |
| France (FR) | 1 | 2 | 18,75 | -1,386E-01 | -5,185E-01 | 2,412E-01 | 43,75 | -5,754E-02 | -3,062E-01 | 1,911E-01 |

**Table 1 Result of Classical and Bayesian two sample tests**

The Bayesian confidence intervals are HPD (High Probability Distribution) intervals. The intervals have been computed with coverage of 90%.

It can be seen that with the present data the classical and the Bayesian point estimator of $\beta$ show the same tendency regarding improvement (positive b) or deterioration (negative $\beta$). In none of the three cases the change is significant.

Furthermore it can be seen that the Fisher information of the estimator for $\beta$ is larger for the Bayes estimator than for the classical one. This is especially evident for small sample sizes. That means that a Bayes estimator can influence the result and this influence is stronger the more information is introduced into the posteriori distribution via the prior distribution.

In the following table we have provided an example with the country "Ex1". Here, the classical statistics would see a significant change (deterioration), whereas the Bayes statistics would see no significant change.

A general tendency can be summarised:

In Bayesian statistics the prior distribution can influence the result, the more, the higher the information contained in the prior is compared with the sample. In these cases, an inappropriately chosen prior might corrupt the result. This is a general result and does not hold true only for a specific statistical model, see e.g. Zellner (1971) chapter 2.11 or Schäbe (1993).

|      | 2005-2009 | 2010-2014 | Classical Test | | | | Bayes Test | | | |
|------|-----------|-----------|----------------|------------------|----------------------|----------------------|-----------------|------------------|----------------------|----------------------|
|      |           |           | Infor-mation I | Beta (esti-mator)| lower confidence level| upper confidence level| Infor-mation I | Beta (esti-mator)| lower confidence level| upper confidence level|
|      |           |           |                |                  |                      |                      |                 |                  |                      |                      |
| Ex1  | 84        | 100       | 1150           | -3,487E-02       | -8,337E-02           | 1,363E-02            | 1175            | -3,413E-02       | -8,211E-02           | 1,386E-02            |

**Table 2 Result with a hypothetical country**

**Conclusion**

In this paper we have shown how railway safety behavior can be modeled with the help of a Poisson distribution connected with the Cox model. We have set up the model for different situations, with equally spaced time intervals, time intervals of different length and developed estimators and tests. This has been done for classical statistics as well as for Bayesian statistics.

We have also considered, how a Bayesian prior can influence the result, especially with small sample sizes. This becomes even more important if not only the safety performance of countries shall be assessed but that of single operators (European Union Agency for Railways (2020)).

Nevertheless, statistics on railway safety behavior should be based on characteristics giving larger sample size as incident statistics rather than on accident statistics, see Braband and Schäbe (2013a).

**References**

Andrasik, R., (2020) Evaluation of safety targets based on Bayesian inference

Braband, J. Schäbe, H., (2013a) Assessment of national reference values for railway safety: a statistical treatment, Journal of Risk and Reliability, 227(4) 405–410, 2013

Braband, J. ,Schäbe, H., (2013b), Comparison of compound Poisson processes as a general approach towards efficient evaluation of railway safety, Proceedings of the ESREL 2013, Safety, Reliability and Risk Analysis: Beyond the Horizon – Steenbergen et al. (Eds) 2014 Taylor & Francis Group, London, ISBN 978-1-138-00123-7), pp. 1297-1302.

Cox D.R., Oakes, D. (1984), Analysis of Survival Date, London, New York, Chapman and Hall

European Union Agency for Railways: Report on Railway Safety and Interoperability in the EU, 2018

European Union Agency for Railways: COMMISSION DELEGATED REGULATION (EU) establishing common safety methods for assessing the safety level and the safety performance of railway operators at national and Union level, Draft, July 2020

EU: COMMISSION DECISION of 5 June 2009 on the adoption of a common safety method for assessment of achievement of safety targets, as referred to in Article 6 of Directive 2004/49/EC of the European Parliament and of the Council (2009/460/EC), 2009

Evans, A. W., (2020), FATAL TRAIN ACCIDENTS ON EUROPE'S RAILWAYS: 1980-2019, Centre for Transport Studies, Department of Civil and Environmental Engineering, Imperial College London, London SW7 2AZ, July 2018.

Schäbe, H. (1993), Post-Data Evaluation of Prior Assessment, Statist. Papers 34 (1993), 339-361.

Zellner, A., (1971), An Introduction to Bayesian Inference in Econometrics, John Wiley and Sons, New York, London, Sydney, Toronto